\documentclass[11pt,sort&compress,preprint]{elsarticle}
\journal{Computer Physics Communications}
\pdfoutput=1

\usepackage{bm}
\newcommand{\ve}[1]{\bm{#1}}

\newcommand{\bmm}{\ve{m}}

\newcommand{\bu}{\ve{u}}

\newcommand{\bp}{\ve{p}}

\newcommand{\bq}{\ve{q}}

\newcommand{\bx}{\ve{x}}

\newcommand{\bs}{\ve{s}}

\newcommand{\bR}{\ve{R}}

\newcommand{\bX}{\ve{X}}

\newcommand{\bF}{\ve{F}}
\newcommand{\bI}{\ve{I}}
\newcommand{\bT}{\ve{T}}
\newcommand{\bD}{\ve{D}}

\newcommand{\bphi}{\boldsymbol{\phi}}
\newcommand{\bxi}{\boldsymbol{\xi}}

\newcommand{\eps}{\varepsilon}

\newcommand\kdivu{K \nabla \cdot \bu}

\newcommand{\bL}{\ve{L}}
\newcommand{\bh}{\ve{h}}

\def\onedot{$\mathsurround0pt\ldotp$}
\def\onedot{$\mathsurround0pt\ldotp$}
\def\cdddot#1{
  \mathbin{\vcenter{\baselineskip.67ex
    \hbox{\onedot}\hbox{\onedot}\hbox{\onedot}%
  }}%
}

\newcommand{\overbar}[1]{\mkern 1.5mu\overline{\mkern-1.5mu#1\mkern-1.5mu}\mkern 1.5mu}

\DeclareTextCommand{\textunderscore}{OT1}{\leavevmode\vbox{\hrule width.6em}}

\newcommand{\dd}{\text{d}}

\usepackage{epsf}
\usepackage{amsmath}
\usepackage{mathtools}
\usepackage{amssymb}
\usepackage{stmaryrd}
\usepackage{setspace}
\usepackage{enumerate}
\usepackage{graphicx}
\usepackage{float}
\usepackage{footnote}
\usepackage{microtype}
\usepackage{scalefnt}
\usepackage{microtype}
\usepackage{xfrac}
\usepackage{transparent}
\usepackage{rotate}
\usepackage{array}
\usepackage{tabu}
\usepackage[normalsize]{subfigure}
\usepackage[mediumspace,mediumqspace,Grey,squaren]{SIunits}
\usepackage{xcolor}
\usepackage{multirow}
\usepackage[margin=1in]{geometry}

\DeclareFontFamily{U}{mathx}{\hyphenchar\font45}
\DeclareFontShape{U}{mathx}{m}{n}{<-> mathx10}{}
\DeclareSymbolFont{mathx}{U}{mathx}{m}{n}
\DeclareMathAccent{\widebar}{0}{mathx}{"73}

\usepackage{tabularx}
\newcolumntype{L}{>{\arraybackslash}X} 
\newcommand{\PreserveBackslash}[1]{\let\temp=\\#1\let\\=\temp}
\newcolumntype{C}[1]{>{\PreserveBackslash\centering}p{#1}}
\newcolumntype{R}[1]{>{\PreserveBackslash\raggedleft}p{#1}}
\newcolumntype{Q}[1]{>{\PreserveBackslash\raggedright}p{#1}}

\begin{document}

\begin{frontmatter}

\title{MFC: An open-source high-order multi-component, \\ multi-phase, and multi-scale compressible flow solver}

\author[add1]{Spencer H. Bryngelson}
\author[add1]{Kevin Schmidmayer}
\author[add2]{Vedran Coralic}
\author[add3]{Jomela C. Meng}
\author[add4]{Kazuki Maeda}
\author[add1]{Tim Colonius}
\ead{colonius@caltech.edu}

\address[add1]{Division of Engineering and Applied Science, 
    California Institute of Technology,
    Pasadena, CA 91125, USA}
\address[add2]{Prime Air, Amazon Inc, Seattle, WA 98108, USA}
\address[add3]{Bosch Research and Technology Center,
Sunnyvale, CA 94085, USA}
\address[add4]{Department of Mechanical Engineering, University of Washington, 
Seattle, WA 98195, USA}

\begin{abstract}

        MFC is an open-source tool for solving multi-component,
        multi-phase, and bubbly compressible flows. It is capable of
        efficiently solving a wide range of flows, including droplet
        atomization, shock--bubble interaction, and gas bubble cavitation. We
        present the 5- and 6-equation thermodynamically-consistent
        diffuse-interface models we use to handle such flows, which are coupled
        to high-order interface-capturing methods, HLL-type Riemann solvers,
        and TVD time-integration schemes that are capable of simulating
        unsteady flows with strong shocks. The numerical methods are
        implemented in a flexible, modular framework that is amenable to future
        development. The methods we employ are validated via comparisons to
        experimental results for shock--bubble, shock--droplet, and
        shock--water-cylinder interaction problems and verified to be free of
        spurious oscillations for material-interface advection and gas--liquid
        Riemann problems. For smooth solutions, such as the advection of an
        isentropic vortex, the methods are verified to be high-order accurate.
        Illustrative examples involving shock--bubble-vessel-wall and
        acoustic--bubble-net interactions are used to demonstrate the full
        capabilities of MFC.

\end{abstract}


\end{frontmatter}

\section*{Program summary}\label{s:summary}

\noindent{\it Title of program}: MFC (Multi-component Flow Code) \\
{\it Licensing provisions}: GNU General Public License 3 \\
{\it Permanent link to code/repository}: \texttt{https://mfc-caltech.github.io} \\
{\it Operating systems under which the program has been tested}: UNIX, macOS, Windows \\
{\it Programming language used}: Fortran 90 and Python \\
{\it Nature of problem}: Computer simulation of multi-component flows
requires careful physical model selection and sophisticated treatment of spatial and 
temporal derivatives to keep solutions both thermodynamically 
consistent and free of spurious oscillations.
Further, such methods should be high-order accurate for  
smooth solutions to reduce computational cost and promote 
sharper interfaces for discontinuous ones.
These problems are particularly challenging for flows with
material interfaces, which are important in numerous
applications. \\
{\it Solution method}: The present software incorporates multiple
physical models and numerical schemes for treatment of compressible 
multi-phase and multi-component flows. Additional physical effects and 
sub-grid models are included, such as an ensemble-averaged bubbly flow model.
The architecture was designed to ensure that further development is straightforward.

\section{Introduction}\label{s:intro}

The multi-component flow code (MFC) is an open source high-order solver for
multi-phase and multi-component flows.  Such flows are central to a wide range
of engineering problems.  For example, cavitating flow phenomena are of
critical importance to the development of artificial heart valves and
pumps~\citep{brennen15}, minimizing injury due to blast
trauma~\citep{laksari15,proud15}, and improving shock- and burst-wave
lithotripsy treatments~\citep{coleman87,pishchalnikov03,ikeda06}.  Bubble
cavitation is also pervasive in flows around hydrofoils, submarines, and
high-velocity projectiles~\citep{saurel08,petitpas09,pelanti14}, during
underwater explosions~\citep{etter13,kedrinskii76}, and within pipe systems and
hydraulic machinery~\citep{streeter83,weyler71}. Unfortunately, cavitation in
these settings is usually detrimental, causing noise and material
deterioration~\citep{weyler71,streeter83}. Other cases of interest include the
breakup of liquid droplets and jets~\citep{meng18,meng15,schmidmayer17},
erosion of aircraft surfaces during supersonic flight~\citep{engel58,joseph99},
shock-wave attenuation of nuclear blasts~\citep{chauvin15}, and needle-free
injection for drug delivery~\citep{tagawa13,veilleux19}.

Robust simulation of multi-component compressible flow phenomena requires a
numerical method that maintains discrete conservation, suppresses oscillations
near discontinuities, and preserves numerical stability. For such simulations
to be computationally efficient, the method used should also be high-order
accurate away from discontinuities. Schemes that can potentially achieve these
requirements can be classified as either interface-tracking or
interface-capturing~\citep{fuster18a}.  Examples of interface-tracking methods
include free-Lagrange~\citep{ball10,turangan17},
front-tracking~\citep{glimm03,cocchi97}, and level-set/ghost fluid
schemes~\citep{abgrall01,liu03,liu11,pan18a,hu06,han14,chang13}; these methods
differ from interface-capturing as they treat material interfaces as sharp
features in the flow. This allows interfacial fluids to have differing
equations of state and ensures that interfacial physics are straightforward to
implement. Unfortunately, they do not naturally enforce discrete or total
conservation, though there have been some recent attempts to partially include
these properties~\citep{denner18,fuster18b}.  Interface-capturing methods
instead treat interfaces as discontinuities in material properties via advected
volume fractions. Such methods are generally more efficient than
interface-tracking schemes~\citep{mirjalili18} and can achieve discrete
conservation by simply solving the governing equations in conservative form.
However, they also smear material interfaces via numerical
diffusion~\citep{johnsen12,perigaud05,shyue99}. In these smeared regions, we
must ensure that the mixture properties are treated in a thermodynamically- and
numerically-consistent way, such that spurious oscillations (and other physical
inconsistencies) are avoided in the presence of high density contrasts between
materials. Fortunately, such methods have been developed and prove to be a
robust treatment of interfacial dynamics~\citep{johnsen06,johnsen12,coralic14}.
Still, the lack of a coherent material interface means that interfacial physics
are more challenging to implement than interface-tracking methods, though
conservative treatments do exist for both interfacial heat
transfer~\citep{massoni02} and capillary
effects~\citep{perigaud05,meng16,schmidmayer17}.

Here, we choose to use an interface-capturing scheme because computational
efficiency and discrete conservation are of principal importance to many
problems of interest. The interface-capturing schemes we use follow from the
so-called 5- and 6-equation models, which are known to be sufficient for
representing a wide range of flow
phenomenologies~\citep{massoni02,allaire02,kapila01,saurel09}.  They are
complemented by a discretely-conservative numerical method that solves the
conservative form of the compressible flow equations. To maintain a
non-oscillatory behavior near material interfaces, the material advection
equations are formulated in a quasi-conservative form~\citep{abgrall96}.  The
equations of motion are then closed by a thermodynamically consistent set of
mixture rules~\citep{allaire02}. The governing equations are solved with a
shock-capturing finite-volume method. Specifically, we adopt a WENO spatial
reconstruction that can achieve high-order accuracy while maintaining
non-oscillatory behavior near material
interfaces~\citep{jiang96,balsara00,henrick05,coralic14}. This scheme is then
coupled with an HLL-type approximate Riemann solver~\citep{toro09,toro94} and a
total-variation-diminishing (TVD) time stepper~\citep{gottlieb98}. 

MFC is, of course, not the only viable option for simulation of compressible
multi-phase flows. For example, ECOGEN~\citep{schmidmayer18} offers a
pressure-disequilibrium-based interface-capturing scheme that is well suited
for the same problems reached by MFC. However, ECOGEN is built upon an
intrinsically low-order MUSCL scheme that can inhibit both efficient simulation
and physically fidelity when compared to the WENO schemes we use here,
especially when augmented with our high-order cell-average approximations. This
is demonstrated in section~\ref{s:vortices} for an isentropic vortex problem
and in section~\ref{s:cavitation} and~\citet{schmidmayer19} for cavitating gas
bubbles. In pursuit of this, MFC was also constructed with a phase-averaged
flow model that represent unresolved multi-phase dynamics at the sub-grid
level. More mature CFD solvers such as OpenFOAM are also available. OpenFOAM
natively supports finite-volume methods for multi-component
flows~\citep{weller98}, though higher-order methods and interface-capturing
models are only available via links to external forked projects. MFC instead
offers an integrated approach that avoids any conflicts from such libraries.
Finally, the parallel I/O file systems we employ ensures that the MFC
architecture can scale up to the largest modern HPC systems.

Herein, we describe version~1.0 of MFC. In section~\ref{s:overview} we
present an overview of what is included in the MFC package, including its
organization, features, and logistics. The physical models we use are presented
in section~\ref{s:model}, including the associated mixture rules, governing
equations, equations of state, and our implementation of an ensemble-averaged
bubbly flow model. The numerical methods used to solve the associated equations
are presented in section~\ref{s:numerics}. A series of test cases simulated
using MFC are discussed in section~\ref{s:examples}; these verify and
validate our method. These are complemented by a set of illustrative example
cases that further demonstrate MFC's capabilities in
section~\ref{s:illustrations} and parallel benchmarking in
section~\ref{s:performance}, which analyzes the performance of MFC on large
scale computing clusters. Section~\ref{s:conclusions} concludes our
presentation of MFC.

\section{Overview and features}\label{s:overview}

\subsection{Package, installation, and testing}\label{s:package}

MFC is available at \texttt{https://mfc-caltech.github.io}.  The source
code is written using Fortran 90 with MPI bindings used for parallel
communication and Python scripts are used to generate input files.
Installation of MFC requires the FFTW package for cylindrical coordinate
treatment~\citep{FFTW98}, and, optionally, Silo and its dependencies for
post-treatment of data files~\citep{millersilo}.  We only provide the FFTW
package with MFC to keep the package relatively small.

\begin{table}
	\centering
	\begin{tabular}{ |           l | l              | }
		\hline 
		\textbf{Name} 	& 	\textbf{Description} \\ \hline
		example\_cases/ 	&	Example and demonstration cases	        \\
		doc/ 			&	Documentation 	                        \\
		installers/ 		&	Package installers: Includes FFTW       \\ 
		lib/ 			&	Libraries                               \\ 
		src/ 			&	Source code                             \\ 
		\rule{1.25cm}{0.3pt}	master\_scripts/ 	&	Python modules and dictionaries, optional source files 	\\ 
		\rule{0.8cm}{0.3pt}	pre\_process\_code/ 	&	Generates initial conditions and grids 	\\ 
		\rule{1.0cm}{0.3pt}	simulation\_code/ 	&	Flow solver 	                        \\ 
		\rule{0.6cm}{0.3pt}	post\_process\_code/	&	Processes simulation data 	        \\ 
		tests/ 		&	Test cases to ensure software is operating as intended 	                \\ 
		AUTHORS	& 	List of contributors and their contact information      \\
		CONFIGURE	& 	Package configuration guide                     \\
		COPYRIGHT	& 	Copyright notice                                \\
		INSTALL		& 	Installation guide                              \\
		LICENSE		& 	The GNU public license file                     \\
		Makefile.in 	& 	Makefile input; generally does not need to be modified 	    \\
		Makefile.user	&  	User inputs for compilation; requires attention from user   \\
        Makefile	&	Targets: all (default), [component], test, clean   \\
		RELEASE		& 	Release notes                                    \\
		\hline
	\end{tabular}
    \caption{Descriptions of the files and directories included in MFC. [component] is
    one of pre\_process, simulation, or post\_process.}
	\label{t:directories}
\end{table}

The MFC package includes several directories; their organization and
descriptions are shown in table~\ref{t:directories}.  New users should consult
the \texttt{CONFIGURE} and \texttt{INSTALL} files for instructions on how to
compile MFC.  In brief, the user must ensure that Python and an MPI Fortran
compiler are loaded and that the FFTW package can be located by
\texttt{Makefile}; this can be done by pointing \texttt{Makefile.user} to the
correct location or by installing the included FFTW distribution in the
\texttt{installers} directory.  Once the software has been built, the
\texttt{test} target of \texttt{Makefile} should be called; it runs multiple
tests (which are located in the tests directory) to ensure that MFC is
operating as intended.

\subsection{Features}\label{s:features}

MFC uses a fully parallel environment via message passing interfaces (MPI),
the performance of which is the subject of section~\ref{s:performance}.
Computationally, it includes structured Cartesian and cylindrical grids with
non-uniform mesh stretching available; characteristic-based Thompson, periodic,
and free-slip boundary conditions have also been implemented. The 5- and
6-equation flow models can be used with a flexible number of components, as
discussed in section~\ref{s:model}. Ensemble-averaged dilute bubbly flow
modeling is also available, including options for Gilmore and Keller--Miksis
single-bubble models (see section~\ref{s:bubbles}). The numerical methods are
discussed in section~\ref{s:numerics}; they include 1-, 3-, and 5-th order
accurate WENO reconstructions on optionally the primitive, conservative, or
characteristic variables. Within each finite volume, high-order evaluation is
available for the cell-averaged variables via Gaussian quadrature for
multi-dimensional problems. The shock-capturing schemes are paired with either
HLL, HLLC, or exact Riemann solvers. For time-stepping, 1--5-th order accurate
Runge--Kutta methods are available.  These features will, of course, evolve and
expand with time. 

Of great practical importance are the user interfaces we utilize. MFC features
Python input scripts, which operate via dictionaries to automatically write
input files that are read via Fortran namelists. Additionally, the file system
and data formats were selected to enable large-scale parallel simulations.
Specifically, we use the Lustre file system to generate and read restart files;
it can support gigabyte-per-second-scale IO operations and petabyte-scale
storage requirements~\citep{halbwachs91}, which ensures that the MFC can
utilize the full capabilities of modern HPC systems. We also utilize HDF5 Silo
databases, which keeps the file structure compact and enables parallel
visualization.  We have confirmed that MFC works as expected on various
high-performance computing platforms, including modern SGI- and Dell-based
supercomputers.

\subsection{Software structure}\label{s:structure}

\subsubsection{Pre-processing}\label{s:preprocess}

The pre-processor generates initial conditions and spatial grids from the
physical patches specified in the Python input file and exports them as binary
files to be read by the simulator. Specifically, this involves allocating
and writing either a Cartesian or cylindrical mesh, with the option of mesh
stretching, according to the input parameters. The specified physical variables
for each patch are transformed into their conservative form and written in a
manner consistent with the mesh. The pre-processor is comprised of individual
Fortran modules that read input values and export mesh and initial condition
files, assign then distribute global variables via MPI, perform variable
transformations, generate grids, parse and assign patch types, and check that
specified input variables are physically consistent and that specified options
do not contradict each other. 

\subsubsection{Simulation}\label{s:simulation}

The simulator, given the initial-condition files generated by the
pre-processor, solves the corresponding governing flow equations with the
specified boundary conditions using our interface-capturing numerical method.
Simulations are conducted for the number of time steps indicated. The
simulator exports run-time information, restart files that can be used to
either restart the simulation or post-process the associated data, and,
optionally, human-readable output data. The structure of the simulator
follows that of pre-processor, with individual Fortran modules conducting
each software component; this includes reading and exporting data and grid
files, performing Fourier transforms, assigning and distributing global
variables via MPI, performing variable transformations, computing time and
spatial derivatives using WENO and the Riemann solver specified, computing
boundary values, including ensemble-averaged bubbly flow physics, and checking
that the input variables are valid.

\subsubsection{Post-processing}\label{s:postprocess}

The post-processor reads simulation data and exports HDF5/Silo databases
that include variables and derived variables, as specified in the input file.
Since the simulator can export human-readable data, post-processing is
not essential for the usage of MFC, but is a useful tool, especially for large
or parallel data structures. Specifically, the post-process component of the
MFC reads the restart files exported by the simulator at distinct time
intervals and computes the necessary derived quantities. The HDF5 database is
then generated and exported, and can be readily viewed using, for example,
VisIt~\citep{visit} or Paraview~\citep{paraview}. Again, individual Fortran
modules perform the associated tasks, including reading data, parameter
conversion, assigning and distributing global MPI variables, computing Fourier
transforms, exporting HDF5 Silo databases, and checking that the input
parameters are consistent.

\subsection{Description of input/output files}\label{s:IO}

\begin{table}
	\centering
	\begin{tabular}{ | c | |           l | l              | }
		\hline 
		& \textbf{Name} & 	\textbf{Description} \\ \hline
		\parbox[t]{3mm}{\multirow{4}{*}{\rotatebox[origin=c]{90}{Input}}} 
		& input.py		&	Input parameters \\
		& \rule{0.8cm}{0.3pt}	pre\_process.inp 	&	Pre-process input parameters, auto-generated \\ 
		& \rule{1.cm}{0.3pt}	simulation.inp 		&	Simulation input parameters, auto-generated \\ 
		& \rule{0.6cm}{0.3pt}	post\_process.inp 	&	Post-process input parameters, auto-generated \\ 	\hline
		\parbox[t]{3mm}{\multirow{6}{*}{\rotatebox[origin=c]{90}{Output}}} 
		& run\_time.inf	& 	Run-time information including current simulation time and CFL \\ 
		& D/			&	Formatted simulation output files  \\
		& p\_all/		&	Binary simulation restart files (depending upon options used) \\
		& restart\_data/	&	Lustre restart files (depending upon options used) \\
		& silo\_HDF5/	&	Silo post-process files (depending upon options used)  \\
		& binary/		&	Binary post-process files (depending upon options used) \\
		\hline
	\end{tabular}
	\caption{Input/output files in the case-specific directory.}
	\label{t:IO}
\end{table}

We next describe the contents of a case-specific directory and its logistics.
The specific file structure is shown in table~\ref{t:IO}.  The Python script
\texttt{input.py} is used to generate the input files
(\textasteriskcentered.inp) for the source codes and execute an MFC component
(one of pre\_process, simulation, or post\_process) either in the active window
or as a submitted batch script. This Python file contains the input parameters
available for the MFC. MFC, depending upon the component used and options
selected, will generate several files and directories. If enabled, the
\texttt{run\_time.inf} file will be generated by the simulator and includes
details about the current time step, simulation time, and stability criterion.
Directories that contain binary restart data and output files for visualization
or further post-treatment are generated, again depending upon the specified
options, by the simulator and post-processor.

\section{Physical model and governing equations}\label{s:model}

The mechanical-equilibrium compressible multi-component flow models we
use can be written as
\begin{gather}
    \frac{\partial \bq}{\partial t} + \nabla \cdot \bF 
    \left( \bq \right) + \bh \left( \bq \right) 
    \nabla \cdot \bu = 
    \bs \left( \bq \right),
    \label{e:goveq}
\end{gather}
where $\bq$ is the state vector, $\bF$ is the flux tensor, $\bu$ is the
velocity field, and $\bh$ and $\bs$ are non-conservative quantities we describe
subsequently.

\subsection{5-equation model}\label{s:5eqn}

We first introduce our implementation of the thermodynamically consistent
mechanical-equilibrium model of~\citet{kapila01}.  Our multi-component
implementation can be used for $N_k$ components, though we present a
two-component ($N_k = 2$) configuration here for demonstration purposes. It
consists of five partial differential equations as 
\begin{gather}
    \bq = \left[ \begin{array}{c}
        \alpha_1 \\
        \alpha_1 \rho_1 \\
        \alpha_2 \rho_2 \\
        \rho \bu \\
        \rho E
    \end{array} \right], \quad
    \bF = \left[ \begin{array}{c}
        \alpha_1 \bu \\
        \alpha_1 \rho_1 \bu \\
        \alpha_2 \rho_2 \bu \\
        \rho \bu \otimes \bu + p \bI - \bT \\
        \left( \rho E + p \right) \bu - \bT \cdot \bu
    \end{array} \right], \quad 
    \bh = \left[ \begin{array}{c}
        - \alpha_1 - K \\
        0 \\
        0 \\
        \mathbf{0} \\
        0
    \end{array} \right], \quad
    \bs =  \mathbf{0} ,
    \label{e:5eqn}
\end{gather}
where $\rho$, $\bu$, and $p$ are the mixture density, velocity, and pressure,
respectively, $\alpha_k$ is the volume fraction of component $k$,
and $\bT$ is the viscous stress tensor
\begin{gather}
    \bT = 2 \eta \left( \bD - \frac{1}{3} (\nabla \cdot \bu) \bI \right), 
    \label{e:viscosity}
\end{gather}    
where $\eta$ is the mixture shear viscosity and
\begin{gather}
    \bD = \frac{1}{2} \left( \nabla \bu + (\nabla \bu)^\top \right)
\end{gather}
is the strain rate tensor.
The mixture total and internal energies are $E = e + \| \bu \|^2/2$ and   
\begin{gather}
    e = \sum_{k=1}^{N_k} Y_k e_k \left( \rho_k , p \right),
    \label{e:ie}
\end{gather}
respectively, where $Y_k = \alpha_k \rho_k / \rho$ are the mass fractions
of each component.
We close~\eqref{e:ie} using the stiffened-gas equation of state, which is chosen
for its ability to faithfully model both liquids and 
gases~\citep{menikoff89}; for component $k$ it is
\begin{gather}
    p_k = ( \gamma_k - 1) \rho_k e_k - \gamma_k \pi_{\infty,k},
    \label{e:EOS}
\end{gather}
where $\gamma$ is the specific heat ratio and $\pi_\infty$ is the 
liquid stiffness (gases have $\pi_\infty = 0$)~\citep{lemetayer04}.
For liquids, these are usually interpreted as fitted parameters
from shockwave-Hugoniot data~\citep{marsh80,gojani09}.
The speed of sound of each component is then
\begin{gather}
	c_k = \sqrt{ \frac{\gamma_k (p_k + \pi_{\infty,k})}{\rho_k} }.
	\label{e:SOS_single}
\end{gather}

The $K$ term in $\bh$ of~\eqref{e:5eqn} represents 
expansion and compression in mixture regions.
For an $N_k = 2$ configuration it is
\begin{gather}
    K = \frac{\rho _2 c_2^2 - \rho _1 c_1^2}{\frac{\rho _2 c_2^2}{\alpha _2} + \frac{\rho _1 c_1^2}{\alpha _1}}
\end{gather}
and the mixture speed of sound $c$ follows as the so-called 
Wood speed of sound~\citep{wood30,wallis69}
\begin{gather}
    \frac{1}{\rho c^2} = \sum_{k=1}^{N_k} \frac{\alpha_k}{\rho_k c_k^2}.
    \label{e:SOS_kdivu}
\end{gather}
Ultimately, the equations are closed by the usual set of mixture rules
\begin{gather}
	1 = \sum_{k=1}^{N_k} \alpha_k, \quad
	\rho = \sum_{k=1}^{N_k} \alpha _k \rho_k, \quad
        \rho e = \sum_{k=1}^{N_k} \alpha_k \rho_k e_k,
        \quad \text{and} \quad
        \eta = \sum_{k=1}^{N_k} \alpha_k \eta_k.
        \label{e:mixture}
\end{gather}

We note that the models of~\citet{allaire02} and~\citet{massoni02} do not
include the $K$ term in~\eqref{e:5eqn} and thus do not strictly obey the
second-law of thermodynamics, nor reproduce the correct mixture speed of
sound~\eqref{e:SOS_kdivu}. While MFC also supports these models, accurately
representing the sound speed is known to be important for some problems, such
as the cavitation of gas bubbles~\citep{schmidmayer19}. However, it is also
known that the $K$ term can result in numerical instabilities for problems with
strong compression or expansion in mixture regions due to its non-conservative
nature~\citep{schmidmayer19}. Thus, the decision of what 5-equation model to
use (if any) is problem dependent and left to the user.

\subsection{6-equation model}\label{s:6eqn}

While the 5-equation model described in section~\ref{s:5eqn} is efficient and
represents the correct physics, the $\kdivu$ term that makes the model
thermodynamically consistent can sometimes introduce numerical
instabilities~\citep{saurel09,schmidmayer19}. In such cases, a pressure-disequilibrium
is preferable~\citep{schmidmayer19}. We also support the
6-equation pressure-disequilibrium model of~\citet{saurel09}, which for a
two-component configuration is expressed as 
\begin{gather}
    \bq = \left[ \begin{array}{c}
        \alpha_1 \\
        \alpha_1 \rho_1 \\
        \alpha_2 \rho_2 \\
        \rho \bu \\
        \alpha _1 \rho _1 e_1 \\
        \alpha _2 \rho _2 e_2
    \end{array} \right], \quad
    \bF = \left[ \begin{array}{c}
        \alpha_1 \bu \\
        \alpha_1 \rho_1 \bu \\
        \alpha_2 \rho_2 \bu \\
        \rho \bu \otimes \bu + p \bI - \bT \\
        \alpha _1 \rho _1 e_1 \bu \\
        \alpha _2 \rho _2 e_2 \bu
    \end{array} \right], \quad 
    \bh = \left[ \begin{array}{c}
        - \alpha_1 \\
        0 \\
        0 \\
        \mathbf{0} \\
        \alpha _1 p_1 \\
        \alpha _2 p_2
    \end{array} \right], \quad
    \bs = \left[ \begin{array}{c}
        \mu \delta p \\
        0 \\
        0 \\
        \mathbf{0} \\
        - \mu p_I \delta p - \alpha_1 \bT_1 : \nabla \bu \\
        \phantom{-} \mu p_I \delta p  - \alpha_2 \bT_2 : \nabla \bu
    \end{array} \right],
    \label{e:6eqn}
\end{gather}
where $\bT_k$ are the component-specific 
viscous stress tensors
and the other terms of $\bs$ represent
the relaxation of pressures between components with 
coefficient $\mu$. The interfacial pressure is 
\begin{gather}
    p_I =\frac{z_2 p_1 + z_1 p_2}{z_1 + z_2},
\end{gather}
where $z_k = \rho _k c_k$ is the acoustic impedance of component $k$ and
\begin{gather}
    \delta p = p_1 - p_2,
\end{gather}
is the pressure difference. Since $p_1 \neq p_2$, 
the total energy equation of the mixture is replaced by the 
internal-energy equation for each component.
The mixture speed of sound is defined according to
\begin{gather}
    c^2 = \sum_{k=1}^2 Y_k c_k^2 ,
    \label{e:SOS_6eqn}
\end{gather}
though after applying the numerical infinite pressure-relaxation procedure detailed in
section~\ref{s:relaxation} the effective mixture speed of 
sound matches~\eqref{e:SOS_kdivu}. 


\subsection{Bubbly flow model}\label{s:bubbles}

\subsubsection{Implementation}

MFC includes support for the ensemble-phase-averaged bubbly flow
model of~\citet{zhang94}, and our implementation of it matches that of~\citet{bryngelson19}.
The bubble population has void fraction $\alpha_b$, which
is assumed to be small, and the carrier components have mixture
pressure $p_l$. The equilibrium radii of the bubble population 
are represented discretely as $\bR_o$, which are $N_\text{bin}$ bins of an assumed
log-normal PDF with standard deviation $\sigma_p$~\citep{colonius08}. The instantaneous
bubble radii are a function of these equilibrium states as 
$\bR(\bR_o) = \{ R_1, R_2, \dots, R_{N_\text{bin}} \}$.
The total mixture pressure is modified as
\begin{gather}
	p = (1-\alpha_b) p_l +
	\alpha_b  \left(
		\frac{\overbar{\bR^3 \bp_{bw} }}{\overbar{ \bR^3}} - 
		\rho \frac{ \overbar{ \bR^3 \dot{\bR}^2 }}{ \overbar{\bR^3} }
	\right),
\end{gather}
where $\dot{\bR}$ are the bubble radial velocities and $\bp_{bw}$ are the bubble wall pressures.
Overbars $\overbar\cdot$ denote the usual moments with respect to the log-normal PDF. 
The bubble void fraction is advected as
\begin{gather}
	\frac{\partial \alpha_b }{\partial t } + \bu \cdot \nabla \alpha_b =
	3 \alpha_b \frac{ \overbar{\bR^2 \dot{\bR} }}{ \overbar{\bR^3} },
\end{gather}
and the bubble dynamic variables are evolved as
\begin{gather}
	\frac{ \partial n \bphi}{\partial t} + \nabla \cdot (n \bphi \bu) = n \dot\bphi,
    \label{e:bdv}
\end{gather}
where $\bphi \equiv \left\{ \bR, \dot\bR, \bp_b, \bmm_v \right\}$ (see section~\ref{s:RPE}) 
and $n$ is the conserved bubble number density per unit volume
\begin{gather}
	n = \frac{3}{ 4 \pi} \frac{\alpha_b}{ \overbar{\bR^3} }.
\end{gather}

\subsubsection{Single-bubble dynamics}\label{s:RPE}

A partial differential equation following~\eqref{e:bdv} 
is evolved for each bin representing equilibrium radius $R_o$.
These equations assume that each bubble evolves, without interaction
with its neighbors, in an otherwise uniform flow whose properties
are dictated by the local mixture-averaged flow quantities~\citep{ando10}.
We also assume that the bubbles remain spherical, maintain a uniform internal
pressure, and do not break-up, or coalesce.
Our model includes the thermal effects, viscous and acoustic damping, and phase change.
The bubble radial accelerations $\ddot{R}$ are computed by the Keller--Miksis equation~\citep{keller80}:
\begin{gather}
	R \ddot{R} \left( 1 - \frac{ \dot{R} }{c_b} \right) + \frac{3}{2} \dot{R}^2 \left( 1 - \frac{ \dot{R} }{3 c_b} \right) =
	\frac{ p_{bw} - p_l }{\rho} \left( 1 + \frac{ \dot{R} }{c_b} \right) + \frac{ R \dot{p}_{bw} }{\rho c_b},
\end{gather}
where $c_b$ is the usual speed of sound associated with the bubble and
\begin{gather}
	p_{bw} = p_b - \frac { 4 \mu \dot{R} } { R } - \frac { 2 \sigma } { R }
\end{gather}
is the bubble wall pressure, 
for which $p_b$ is the internal bubble pressure, $\sigma$ is the surface tension coefficient,
and $\mu$ is the liquid viscosity. The evolution of $p_b$ is evaluated using the model of~\citet{ando10}:
\begin{gather}
	\dot{p}_{b} = \frac { 3 \gamma_b }{R} \left(  
	\mathfrak{R}_{v} T_{bw} \dot{m}_v - 
	\dot{R} p_{b} +
	\frac{ \gamma_b - 1 } { \gamma_b } \lambda_{bw}  \left.  \frac { \partial T } { \partial r } \right|_{r=w}  
	\right),
\end{gather}
where $T$ is the temperature, $\lambda$ is the thermal conductivity,
$\mathfrak{R}_v$ is the gas constant and $\gamma_b$ is the specific heat ratio of the gas.
Mass transfer of the bubble contents follows the reduced model of~\citet{preston07} as
\begin{gather}
		\dot{m}_v = \frac { \mathcal{D} \rho_{bw} } { 1 - \chi_{vw} } \left.
		\frac { \partial \chi_v } { \partial r } \right|_w.
\end{gather}

\section{Solution method}\label{s:numerics}

Our numerical scheme generally follows that of \citet{coralic14}.
The spatial discretization of~\eqref{e:goveq} 
in three-dimensional Cartesian coordinates is
\begin{gather}
	\frac{ \partial \bq }{\partial t} +
	\frac{\bF^{x}(\bq)}{\partial x}  +
	\frac{\bF^{y}(\bq)}{\partial y} +
	\frac{\bF^{z}(\bq)}{\partial z} =
	\bs(\bq) - \bh(\bq) \nabla \cdot \bu,
	\label{e:partial}
\end{gather}
where $\bF^{x_i}$ are the $i \in (x,y,z)$-direction flux vectors
and the treatment of $\nabla \cdot \bu$ is discussed later.

\subsection{Treatment of spatial derivatives}

We use a finite volume method to treat the spatial derivatives
of~\eqref{e:partial}.
The finite volumes are
\begin{gather}
	I_{i,j,k} = 	[ x_{i-1/2}, x_{i+1/2} ] \times  	
			[ y_{j-1/2}, y_{j+1/2} ] \times
			[ z_{k-1/2}, z_{k+1/2} ].
\end{gather}
We spatially integrate~\eqref{e:partial} within each cell-centered
finite volume as
\begin{gather}
\begin{split}
	\frac{ \dd \bq_{i,j,k} }{\dd t} =
	&\frac{1}{\Delta x_i} [ \bF_{i-1/2,j,k}^x - \bF_{i+1/2,j,k}^x ] + \\
	&\frac{1}{\Delta y_j} [ \bF_{i,j-1/2,k}^y - \bF_{i,j+1/2,k}^y ] + \\
	&\frac{1}{\Delta z_k}[ \bF_{i,j,k-1/2}^z - \bF_{i,j,k+1/2}^z ] +
	\bs(\bq_{i,j,k}) - \bh(\bq_{i,j,k}) (\nabla \cdot \bu)_{i,j,k},
\end{split}
\label{e:discrete}
\end{gather}
where  
\begin{gather}
	\bq_{i,j,k} = \frac{1}{V_{i,j,k}} \iiint_{I_{i,j,k}} \bq (x,y,z,t) \, \dd x \dd y \dd z, \\
	\bs_{i,j,k} =  \frac{1}{V_{i,j,k}} \iiint_{I_{i,j,k}} \bs (x,y,z,t) \, \dd x \dd y \dd z, \\
	\bF_{i+1/2,j,k} = \frac{1}{\Delta y_j \Delta z_k} \iint_{A_{i+1/2,j,k}} \bF (x,y,z,t) \, \dd y \dd z,
\end{gather}
are cell-volume and face averages, for which 
$\Delta x_i = x_{i+1/2} - x_{i-1/2}$ are the mesh spacings 
($\Delta y$ and $\Delta z$ have the same form),
and $V_{i,j,k}$ and $A_{i+1/2,j,k}$ are the cell volumes and face areas.

MFC can approximate this equation using high-order quadratures as opposed to simple
cell-centered averages. This approach computes the flux surface integrals and source terms 
using a two-point, fourth-order, Gaussian quadrature rule, e.g.
\begin{gather}
	\bF_{i+1/2,j,k} = \frac{1}{4} \sum_{m=1}^2 \sum_{l=1}^2 \bF(\bq(x_{i+1/2}, y_{j_l}, z_{k_m} )),
	\label{e:quad}
\end{gather}
where $l$ and $m$ are the Gaussian quadrature point indices and 
\begin{gather}
	y_{j_l} = y_j + (2l+1) \frac{\Delta y_j}{2 \sqrt{3}} \quad \text{and} \quad 
	z_{k_m} = z_k + (2m+1) \frac{\Delta z_k}{2 \sqrt{3}}.
\end{gather}
The divergence terms are treated using a midpoint rule 
\begin{gather}
\begin{split}
	(\nabla \cdot &\bu)_{i,j,k} =  	\\
		&\frac{1}{\Delta x_i} (u_{i+1/2,j,k} - u_{i-1/2,j,k}) + 
		\frac{1}{\Delta y_j} (v_{i,j+1/2,k} - v_{i,j-1/2,k}) + 
		\frac{1}{\Delta z_k}(w_{i,j,k+1/2} - w_{i,j,k-1/2}),
\end{split} 
\end{gather}
where $\bu = \{ u,v,w \}$ are the cell-averaged velocity components computed
analogous to~\eqref{e:quad}. 

To avoid spurious oscillations at material interfaces, we ultimately evaluate
the fluxes by reconstructing the primitive or characteristic variables at the
cell faces via a 5th-order-accurate WENO scheme~\citep{coralic14} (though
reconstruction of the conservative variables and lower-order WENO schemes are
also supported).  This allows us to apply a Riemann solver, so
\begin{gather}
	\bF_{i+1/2,j,k} = \frac{1}{4} 
		\sum_{m=1}^2 \sum_{l=1}^2 \widehat{\bF}( \bq_{i+1/2,j,k}^L, \bq_{i+1/2,j,k}^R ),
\end{gather}
where $\widehat{\bF}$ is the numerical flux function of the Riemann solver.
We use the HLLC approximate Riemann solver to compute the 
fluxes~\citep{toro94}, though other Riemann solvers
are also supported. 

\subsection{Limiters for improved numerical stability}\label{s:stability}

\subsubsection{Volume fraction limiting}\label{s:vf}

Following our mixture rules~\eqref{e:mixture}, the volume fractions are
physically required to sum to unity. However,
the accumulation of numerical errors can preclude this if the first $N_k - 1$
volume fractions exceed unity, and thus one of the volume fractions must be
negative~\citep{meng16b}. This is of course unphysical and leads to other
numerical issues, such as complex speeds of sound. To treat this, we impose the
volume fraction mixture rule by limiting each volume fraction as $0 \le
\alpha_k \le 1$ for all components $k$, then rescaling them as
\begin{gather}
    \alpha_k = \frac{\alpha_k}{\sum_{k=1}^{N_k} \alpha_k} \quad \text{for} \;  k = 1,\dots,N_k.
\end{gather}
We note that we only use this limiting when computing the mixture properties,
and do not alter them otherwise as to avoid polluting the mass conservation
properties of the method.

\subsubsection{Flux limiting}\label{s:flux}

The WENO schemes we utilize are, in general, not TVD. This can be problematic
when WENO cannot form a smooth stencil to reconstruct, and can lead to
numerical instabilities even when the usual CFL criteria are met. MFC supports
advective flux limiting to treat this issue, which improves stability, though
it also increases numerical dissipation and thus smearing of material
interfaces. However, we use the gradient of the local volume fraction $\chi$ to
minimize this effect, which localizes the limiter to non-smooth regions of the
flow. In one dimension of our dimensional-splitting procedure this yields
\begin{gather}
    \chi_i = \begin{dcases*}
        \frac{ \alpha_{i} - \alpha_{i-1} }{ \alpha_{i+1} - \alpha_i } & if $u^* \ge 0$, \\
        \frac{ \alpha_{i+2} - \alpha_{i+1} }{ \alpha_{i+1} - \alpha_i } & if $u^* < 0$,
    \end{dcases*}
\end{gather}
where $i$ is the spatial index and $u^*$ is the local velocity computed by the
Riemann solver.  Ultimately, the modified flux is a combination of a low- and
high-order accurate flux approximation. The low-order flux is chosen to be
equivalent to a first-order WENO reconstruction and the high-order flux is the
Riemann flux from the WENO reconstruction.  Specifically, MFC supports the
minmod, MC, ospre, superbee, Sweby, van Albada, or van Leer flux limiters, each
of which is a function of the volume fraction gradient $\chi$.  Further details
of our numerical implementation are located in~\citet{meng16b}.

\subsection{Cylindrical coordinate considerations}\label{s:cylindrical}

MFC also supports the use of cylindrical coordinates. They are formulated in
similar fashion to~\eqref{e:partial}, though with the addition of an additional
set of source terms on the right-hand-side associated with the $1 / r$
cylindrical terms of the divergence operator. Our implementation was detailed
by~\citet{meng16b} and does not use an alternative treatment of the WENO
weights in the azimuthal direction, such as would be required to ensure
high-order accuracy away from discontinuities. While such methods for this
exist, they are generally complex and suffer from numerical stability
issues~\citep{li03,mignone14,wang13}. The cylindrical coordinate treatment also
requires velocity gradients, rather than just velocities, at cell boundaries.
For this, we use a second-order-accurate finite-difference and averaging
procedure to obtain the necessary velocity gradients at these points.  Finally,
the coordinate singularity at $r \to 0$ is treated using the method
of~\citet{mohseni00}, which performs differentiation in the radial direction
via a redefinition of the radial coordinate; in our implementation, we place the
singularity at the finite-volume cell boundary (rather than the center).  We note
that this implementation requires an even number of grid cells in the azimuthal
coordinate direction. A consequence of the cylindrical coordinate system is
that the grid cells near the radial center are much smaller than those far
away, which restricts the global CFL criterion. Following~\citet{mohseni00}, we
address this issue by using a spectral filter, which filters the high-frequency
components of the solution near the centerline and relaxes the CFL criterion.
Our implementation was verified by~\citet{meng16b} to be second-order accurate
away from discontinuities via a propagating spherical pressure pulse, while the
construction of the viscous stress tensor was verified using the method of
manufactured solutions.

\subsection{One-way acoustic wave generation}\label{s:asrc}

The source term $\bs$ of~\eqref{e:goveq} and~\eqref{e:partial} can be augmented
with additional terms $\bs^s (t)$ for the generation of one-way acoustic
waves, for examples to model an ultrasound transducer immersed in the flow.
Following~\citet{maeda17}, these take the form
\begin{gather}
    \bs^s (t) = \int_\Gamma \Omega_\Gamma (\bxi, t) \delta ( \bX(\bxi,t) - \bx ) \dd \bxi
\end{gather}
where $\Omega_\Gamma$ is (possibly time dependent) forcing,
$\Gamma$ is the surface the forcing acts upon,
$\delta$ is the Dirac delta function,
and $\bX$ maps the $\bxi$ coordinate
to physical space $\bx$. In our numerical framework
this is represented at cell $I_{i,j,k}$ as
\begin{gather}
    \bs_{i,j,k}^s (t) = \sum_{m=1}^M \Omega_\Gamma (\xi_m,t) \delta_h ( | \bX(\xi_m ,t) - \bx_{i,j,k} | ) \Delta \xi_m,
\end{gather}
where $\delta_h$ is the discrete delta function operator, $\Delta \xi_k$ 
are the sizes of the discrete patch or line the forcing is applied to, and $M$ are the
number of such patches. For example, in two dimensions the 
one-dimensional line operator follows as
\begin{gather}
    \delta_h (h) = \frac{1}{2 \pi \sigma^2} e^{- \frac{1}{2} \frac{h^2}{\sigma^2}},
\end{gather}
where $\sigma = 3 \Delta$ and $\Delta$ is the largest mesh spacing.
For a single-component, two-dimensional problem (note that there
is no volume-fraction advection equation in this case) the forcing
is expressed as 
\begin{gather}
    \Omega_\Gamma (t) = f(t) [ 1 / c_o, \cos(\theta), \sin(\theta), c_o^2 / (\gamma - 1) ], 
\end{gather}
where $f(t)$ is the time-dependent pulse amplitude, $c_o$ is the speed of sound,
and $\theta$ is the forcing direction as measured from the first-coordinate
axis. 

\subsection{Time integration}

Once the spatial derivatives have been approximated, \eqref{e:partial}~becomes
a semi-discrete system of ordinary differential equations in time. We treat the
temporal derivative using a Runge--Kutta time-marching scheme for the state
variables. To achieve high-order accuracy and avoid spurious
oscillations, we use the third-order-accurate total variation diminishing
scheme of~\citet{gottlieb98}:
\begin{align}
	\bq_{i,j,k}^{(1)} &= \bq_{i,j,k}^n + 
            \Delta t \bL(\bq_{i,j,k}^n), \nonumber\\
	\bq_{i,j,k}^{(2)} &= \frac{3}{4} \bq_{i,j,k}^n + 
            \frac{1}{4} \bq_{i,j,k}^{(1)} + 
            \frac{1}{4} \Delta t \bL(\bq_{i,j,k}^{(1)}), \label{e:time} \\
	\bq_{i,j,k}^{n+1} &= \frac{1}{3} \bq_{i,j,k}^n + 
            \frac{2}{3} \bq_{i,j,k}^{(2)} + 
            \frac{2}{3} \Delta t \bL(\bq_{i,j,k}^{(2)}), \nonumber
\end{align}
where~$(1)$ and $(2)$ are intermediate time-step stages, $\bL$ represents the
right hand side of~\eqref{e:discrete}, and $n$ is the time-step index. We note
that MFC also supports Runge--Kutta schemes of orders 1--5.

\subsection{Pressure-relaxation procedure}\label{s:relaxation}

The pressure-disequilibrium model~\eqref{e:6eqn} requires a pressure-relaxation
procedure to converge to an equilibrium pressure.  We use the
infinite-relaxation procedure of~\citet{saurel09}. At each time step, it solves
the non-relaxed hyperbolic equations ($\mu \to 0$) using first-order-accurate
explicit time step integration and a re-initialization procedure that ensures
total energy conservation at the discrete level. After this, the disequilibrium
pressures are relaxed as $\mu \to +\infty$. This procedure is performed at each
Runge--Kutta stage, and so there is a unique pressure at the end of each stage
and the 5- and 6-equation models reconstruct the same variables.  As a result,
simulations of the pressure-disequilibrium model are only modestly more
expensive than the 5-equation models (about 5\% for spherical bubble
collapses~\citep{schmidmayer19}).

\section{Simulation verification and validation}\label{s:examples}

We next present several test cases that validate and verify MFC's
capabilities. These include one-, two-, and three-dimensional cases that span a
wide variety of flow problems.

\subsection{Shock--bubble interaction}\label{s:shockbubble}

We first consider a Mach $1.22$ shock wave impinging on a $\unit{5}{\centi\meter}$
diameter spherical helium bubble in air.  This problem was investigated via
experimental methods by~\citet{haas87} and has been previously used as a
validation case for multi-component flow
simulations~\citep{terashima09,fedkiw99,hu04}.  Our simulations are performed
using an axisymmetric configuration; further simulation specifications can be
found in~\citet{coralic14}.

\begin{figure}
	\centering
    \includegraphics[scale=1]{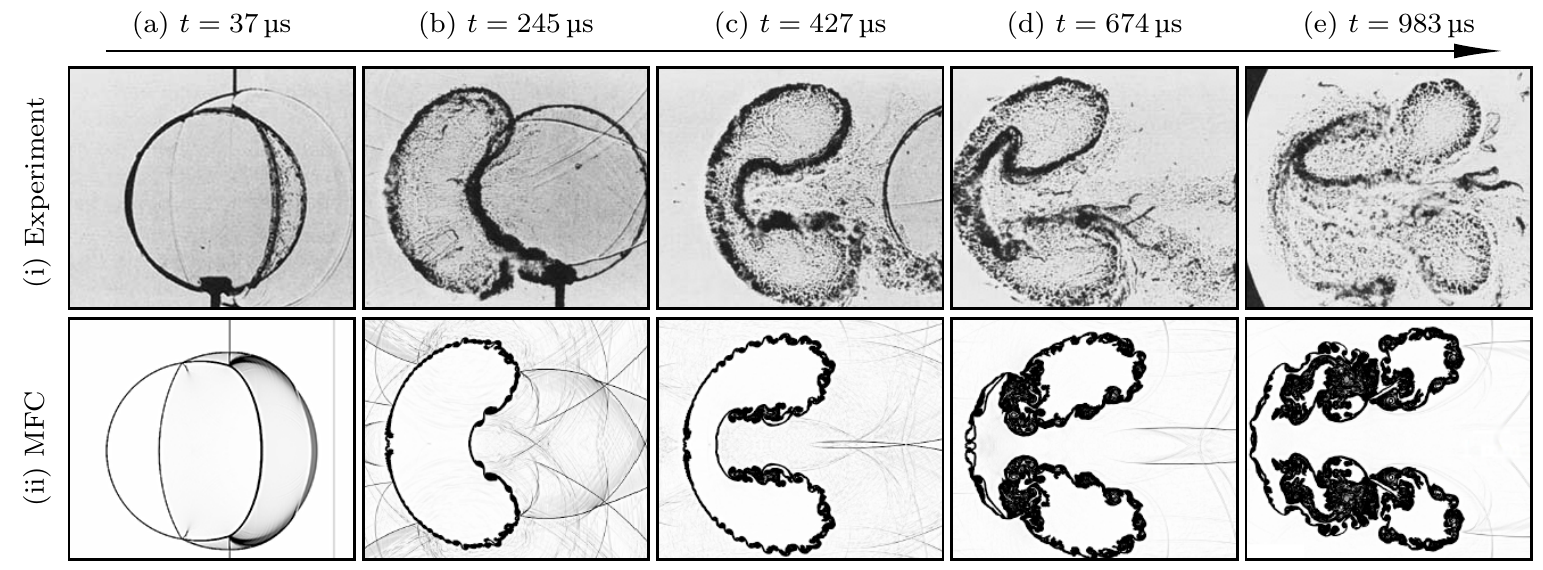}
	\caption{Comparison between (i) experimental shadowgraphs of~\citet{haas87} 
	and (ii) numerical schlieren 
	visualizations~\citep{quirk96} using MFC at 
	select times (a)--(e) as labeled.  
	Experimental images are 
	\textcopyright Cambridge University Press 1987. 
	}
	\label{f:shockbubble}
\end{figure}

Visualizations of the shock impinging the bubble and subsequent breakup and
vortex ring production are shown in figure~\ref{f:shockbubble}.  We see that
the simulation results qualitatively match those of the experiment.
Importantly, no spurious oscillations can be seen in the numerical schlieren
images, despite their sensitivity to small density differences.  We
quantitatively compare our results to the experiment by considering the
velocity of key flow features: \citet{haas87}~measured the velocity of the
incident, reflected, and transmitted shocks, as well as the up and down-stream
interfaces and jets. Our simulation results are within 10\% of the experiments
for all cases and are generally within about 5\% of the experimental means. Our
results are also consistent with those computed independently via the level
set~\citep{hejazialhosseini10} and diffuse-interface methods~\citep{so12},
including the Kelvin--Helmholtz instability that develops along the bubble
interface (see figure~\ref{f:shockbubble}~(b)--(e)).

\subsection{Shock--droplet/cylinder interaction}\label{s:shockdroplet}

We next consider air shocks interacting with liquid media in two
and three dimensions.
These problems are more computationally challenging, primarily due to the
larger density ratio. The first case we analyze consists of a Mach $1.47$ shock
impinging a $\unit{4.8}{\milli\meter}$ diameter liquid water cylinder. The
simulation parameterization can be found in~\citet{meng15}.

\begin{figure}[H]
	\centering
    \includegraphics[scale=1]{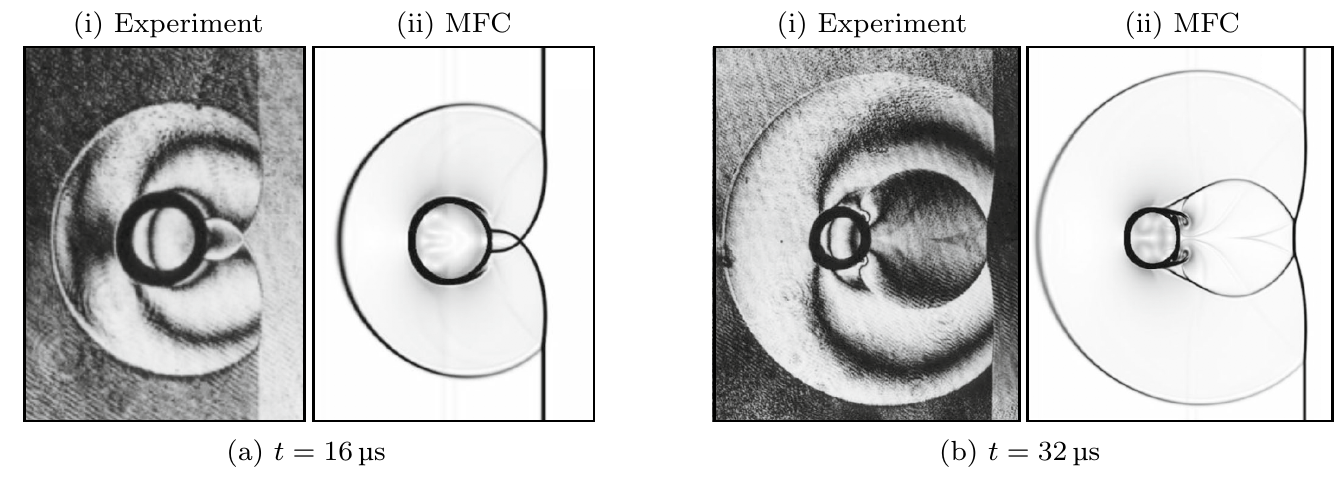}
	\caption{Comparison between (i) holographic interferograms~\citep{igra01} and
	(ii) numerical schlieren visualizations using MFC at select times (a)
	and (b) as labeled. The experimental images are reprinted from \citet{igra01}. }
	\label{f:shockcylinder}
\end{figure}

Figure~\ref{f:shockcylinder} shows a visualization of experimental and our
numerical results as the shock passes over the liquid cylinder.  At early times
it is difficult to assess the cylinder's deformation, so we instead compare the
primary and secondary waves that are generated. In
figure~\ref{f:shockcylinder}~(a) we see that the primary wave system, including
the incident and reflected shock, have the same locations for both experimental
and numerical results. The secondary wave system is generated when the Mach
stems on both sides of the cylinder converge to the rear stagnation point; in
figure~\ref{f:shockcylinder}~(b) we see that these also match closely.

\begin{figure}
	\centering
    \includegraphics[scale=1]{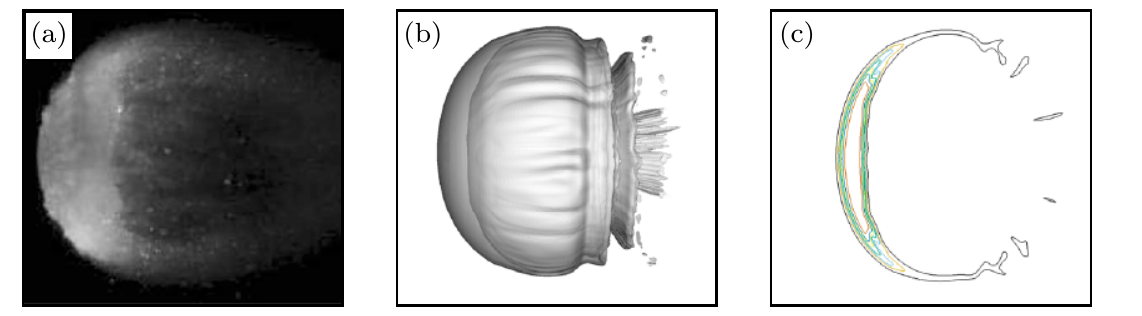}
	\caption{Comparison of (a) experimental~\citep{theofanous12} 
	and (b) numerical water droplets isosurface $\alpha_l = 0.01$.  (c)
	shows isocontours of $\alpha_l$ ranging from $0.01$ to $0.99$. }
	\label{f:shockdroplet}
\end{figure}

We also consider the breakup of a spherical water droplet due to a helium shock
(Mach $0.59$ observed in the post-shock flow), following the experimental
conditions of~\citet{theofanous12}. A full exposition of the simulation
conditions can be found in~\citet{meng18}. Figure~\ref{f:shockdroplet} shows
their experimental image and volume fraction isosurfaces and sliced isocontours
from our simulations.  We use a small $\alpha_l = 0.01$ value for the
isosurface of figure~\ref{f:shockdroplet}~(b) for comparison purposes since
images from experiments are often obscured by the fine mist generated. While it
is challenging to obtain accurate timing data from the experiments, a
qualitative agreement between experiment and simulation are still observed for
the shear-induced entrainment of the droplet.  

\subsection{Spherical bubble dynamics}\label{s:cavitation}

Numerical simulation of cavitating spherical gas bubbles is challenging because
mixture-region compressibility must be properly treated, discrete conservation
must be enforced, and sphericity should be maintained in the presence of large
density and pressure ratios. We consider a collapsing and rebounding air bubble
in water at 10 times higher pressure as a test of the capabilities of MFC.
Specific simulation specifications were presented in~\citet{schmidmayer19};
further, we include a guided description of the simulation setup for this
problem in the \texttt{example\_cases/3D\_sphbubcollapse} directory of the MFC
package (see table~\ref{t:directories}).

\begin{figure}[H]
	\centering
    \includegraphics[scale=1]{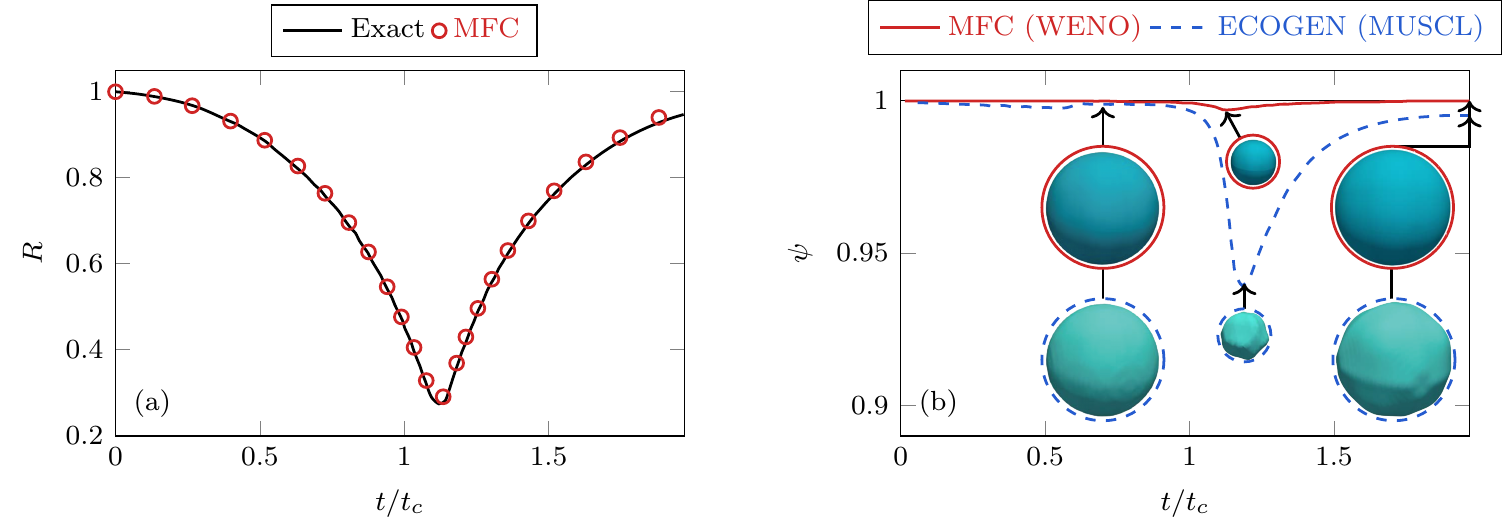}
	\caption{
	Evolution of (a) the dimensionless bubble radius $R$ and (b) its sphericity $\psi$.
    	In (b) the nominal bubble shapes, represented by $\alpha_l = 0.5$ 
        isosurfaces, are shown at select times for MFC (WENO) and ECOGEN (MUSCL).
        }
	\label{f:cavitation}
\end{figure}

Figure~\ref{f:cavitation}~(a) shows the evolution of the bubble radius; it
reaches a minimum near the nominal Rayleigh collapse time
$t_c$~\citep{brennen95}, then rebounds, as expected.  The radius $R$ of our
simulations is computed from the gas volume by assuming the shape is nearly
spherical; this closely matches the solution expected following the
Keller--Miksis equation~\citep{keller80}.  We assess and compare the quality of
our simulation with ECOGEN~\citep{schmidmayer18} via computation of the bubble
sphericity during the collapse-rebound process. Figure~\ref{f:cavitation}~(b)
shows this sphericity $\psi$ of the bubble, which is defined
following~\citet{wadell35} and a value of 1 indicates a spherical bubble. We
see that the WENO scheme used by MFC can better maintain sphericity
than a MUSCL scheme of ECOGEN formulated for the same diffuse-interface
model~\citep{schmidmayer19}, and is thus preferable for this problem.

\subsection{Isentropic and Taylor--Green vortices}\label{s:vortices}

We next consider two- and three-dimension vortex problems as a means to verify
our solutions of the flow equations.  The two-dimension problem we consider is
the evolution of a steady, inviscid, isentropic, ideal-gas vortex. This problem
has been used previously to assess the convergence properties of high-order
WENO schemes for smooth solutions to the Euler
equations~\citep{balsara00,titarev04}.  We use it here to verify that we obtain
high-order accuracy away from shocks and material interfaces; details regarding
the numerical parameters and exact problem formulation can be found
in~\citet{coralic14}.

\begin{figure}[H]
	\centering
    \includegraphics[scale=1]{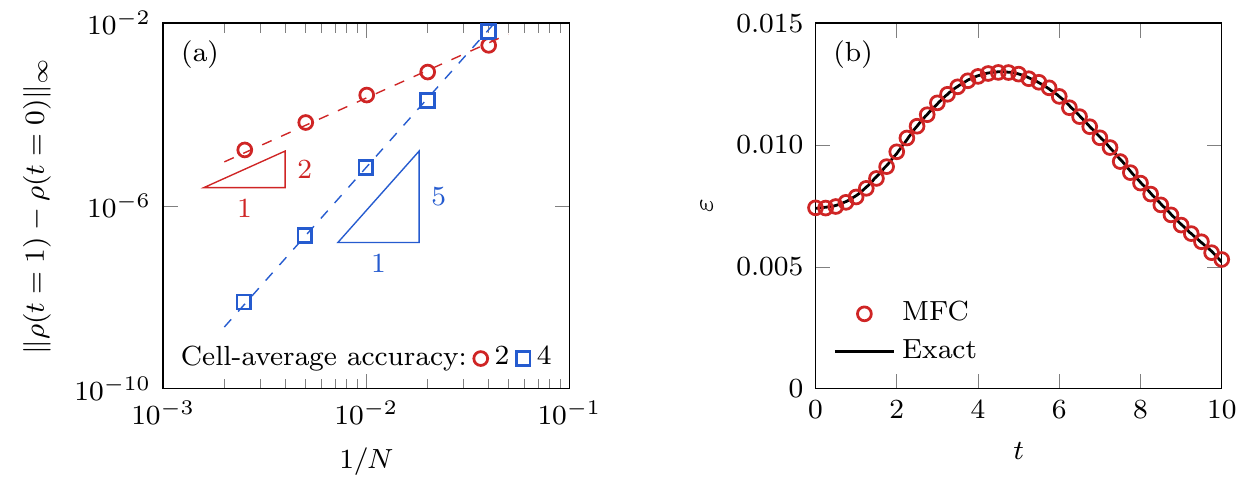}
	\caption{(a) $L_\infty$ density error associated with an isentropic
	steady vortex for the cell-averaged scheme accuracy labeled. 
	(b) Non-dimensional kinetic energy dissipation rate $\eps$ associated
	with the three-dimensional Taylor--Green vortex problem; 
	the MFC solution is compared to the direct spectral solution 
	of~\citet{brachet83}.} 
	\label{f:vortex}
\end{figure}

Figure~\ref{f:vortex}~(a) shows the density error for both low- and high-order
finite volume cell-averaging (following section~\ref{s:numerics}). Since the
solution should be steady, the error is computed as the deviation from the
initial condition after 1 dimensionless time unit as a function of the grid
size, for which $N$ is the number of finite volumes in one spatial direction.
The convergence is 2nd- and 5th-order accurate for 2nd- and 4th-order-accurate
cell-averaging schemes, respectively. Thus, we conclude that for
multi-dimensional problems the 4th-order-accurate cell averaging we employ is
required to achieve 5th-order accuracy associated with the WENO
reconstructions.

We use the three-dimensional Taylor--Green vortex problem of~\citet{brachet83}
to study the production of small length scales, including vortex stretching and
dissipation.  The simulation parameters again follow from~\citet{coralic14}.
Figure~\ref{f:vortex}~(b) shows the dimensionless dissipation rate of the
kinetic energy $\eps$ in dimensionless time $t$, as computed over the entire
computational domain.  We see that the vortex stretching grows until  $t
\approx 5$, after which the effects of viscous dissipation begin to dominate.
MFC results closely match those of the direct solution computed
by~\citet{brachet83} using spectral methods.

\subsection{Further verification}\label{s:further}

MFC has been verified using several other problems, including several
one-dimensional test cases. For example, of great importance are the
development of spurious oscillations at material interfaces, which can
significantly pollute simulation quality. To determine if such oscillations
appear when using our method, \citet{coralic14} considered the advection of an
isolated air--water interface at constant velocity in a periodic domain. It was
shown that when conservative variables are reconstructed, the interface is
corrupted by spurious oscillations and the pressures can even become negative.
Whereas when primitive variables are reconstructed, their character is
oscillation free for both velocity and pressure down to round off error. 

It is also challenging to predict the correct position and speed of waves that
emit from shock--interface interactions. While the shock--bubble interaction
problem considered in section~\ref{s:shockbubble} showed that our method can
approximate these quantities when comparing to experiments, it is helpful to
consider a similar problem in one-dimension where exact solutions are
available. Following~\citet{liu03}, \citet{coralic14} used the methods
employed by MFC to analyze a Mach $8.96$ helium shock wave impinging an air
interface. It was shown that the numerical results quantitatively matches the
associated exact solution, correctly identifying the position and speed of all
waves in the problem while avoiding any spurious oscillations. Of similar
character is the gas--liquid shock tube problem of~\citet{cocchi96}, which has
been used as a model for underwater explosions. For this, \citet{coralic14}
also showed that the numerical solution matches the exact one and correctly
identifies the position and speed of all waves. 

Finally, the ensemble-averaged bubbly flow model introduced in
section~\ref{s:bubbles} was verified by simulating a weak acoustic pulse
impinging a dilute bubble screen and comparing to the linearized bubble dynamic
results of~\citet{commander89}. We saw that the measured phase speed and
acoustic attenuation, computed via the method of~\citet{ando10} and
\citet{bryngelson19}, match the expected results. Further,
\citet{bryngelson19} showed that this method quantitatively matches the
volume-averaged formulation of the same problem~\citep{fuster11,maeda18}.

\section{Illustrative examples}\label{s:illustrations}

\subsection{Shock-bubble dynamics in a vessel phantom}\label{s:vessel}

\begin{figure}[H]
   \centering
   \includegraphics[scale=1]{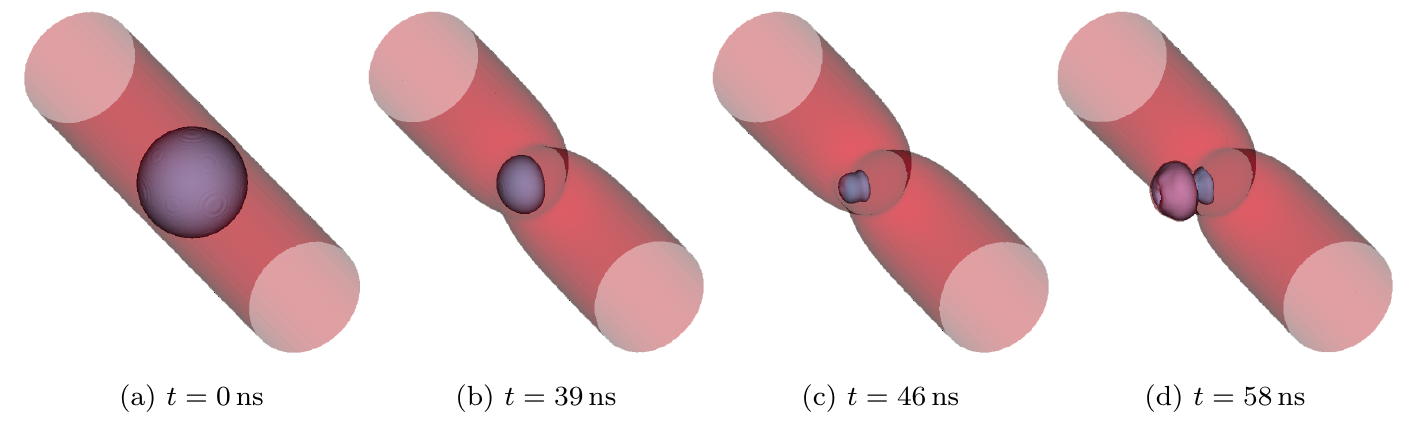}
   \caption{
       Temporal snapshots (a)--(d) that show the bubble collapse and vessel
       wall deformation. The shock impinges the vessel from right to left. 
       The bubble and vessel walls are illustrated via their $0.5$ volume-fraction 
       isosurfaces.
   }
   \label{f:vessel}
\end{figure}

We demonstrate the capabilities of MFC by first considering the
shock-induced collapse of a gas bubble inside a deformable vessel. This is
closely related to the vascular injury that can occur during shock-wave
lithotripsy treatments~\citep{coralic13,coralic14}. This problem is
particularly challenging because it involves lareg density-, pressure-, and
viscosity-ratios as well as components with significantly different equation of
state parameters. Specifically, we consider a $\unit{20}{\micro\meter}$
diameter air bubble centered in a $\unit{26}{\micro\meter}$ diameter
cylindrical vessel filled with water and surrounded by $10\%$ gelatin (see
figure~\ref{f:vessel}~(a)). The problem is initialized via a
$\unit{40}{\mega\pascal}$ shock wave impinging the side of the vessel. Details
on the specific numerical setup can be found in~\citet{coralic15}.

We visualize the bubble dynamics and subsequent impingement and deformation of
the vessel in figure~\ref{f:vessel}. We see that by $t =
\unit{39}{\nano\second}$ the bubble shape is asymmetric and the vessel
contracts due to the incoming shock, after $t = \unit{46}{\nano\second}$ the
bubble surface has gained an inflection point and by $t =
\unit{66}{\nano\second}$ it impinges the vessel surface and becomes
mushroom-shaped. Importantly, all surfaces remain smooth and free of spurious
oscillations, despite the large density- and viscosity-ratios. 

\subsection{Vocalizing humpback whales}\label{s:whales}

We next demonstrate the utility, flexibility, and robustness of the
ensemble-averaged bubbly flow model described in section~\ref{s:bubbles}. For
this, we model the humpback whale bubble-net feeding process~\citep{hain81}.
Specifically, multiple humpback whales vocalize towards an annular bubbly
region called a bubble net, which is modeled using the acoustic source terms of
section~\ref{s:asrc} and bubbly regions described using the phase-averaged
model. 

\begin{figure}[H]
   \centering
   \includegraphics[scale=1]{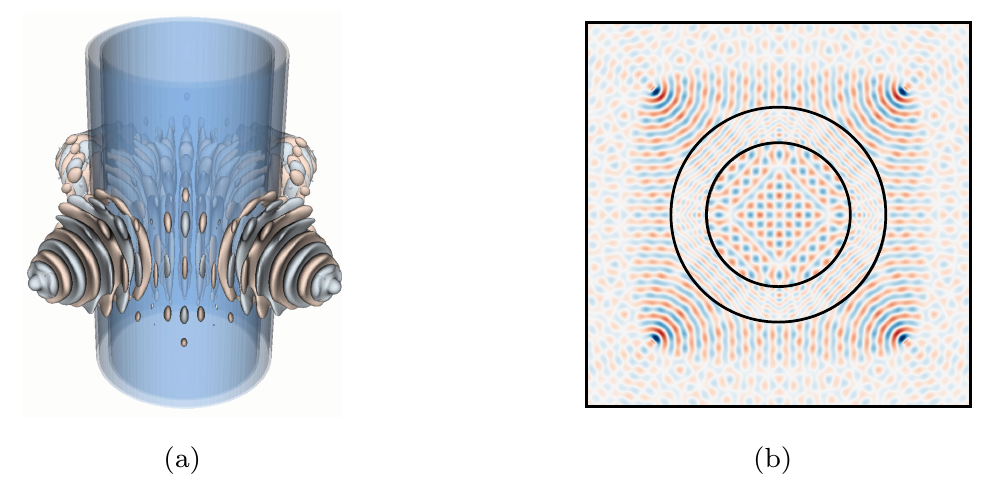}
   \caption{
        Visualizations of four model humpback whales
        vocalizing towards an annular bubble net in (a) three- and (b) two-dimensions.
        The acoustics are shown via isocontours of pressure and the annular
        bubble net is shaded.
   }
   \label{f:nets}
\end{figure}

We show the acoustics associated with the periodic excitation of the bubble net
in figure~\ref{f:nets} for both two- and three-dimensional configurations. In
both cases, we see that the impedance associated with the relatively dilute
bubble net (void fraction $10^{-4}$) effectively shields the core region from
the vocalizations. Indeed, it is anticipated that the whales use their nets as
a tool for corralling their prey into this relatively quiet, compact region. We
also see that the curved material interfaces remain smooth and free of spurious
oscillations.

\section{Parallel performance benchmarks}\label{s:performance}

It is important to ensure that our parallel implementation can utilize modern,
large computer resources. To do this, we benchmark MFC's parallel
architecture via the usual scaling and speedup tests. These tests were carried
out using 288 compute nodes, each containing two $\unit{2.6}{\giga\hertz}$
six-core AMD Opteron processors.

\begin{figure}[H]
	\centering
    \includegraphics[scale=1]{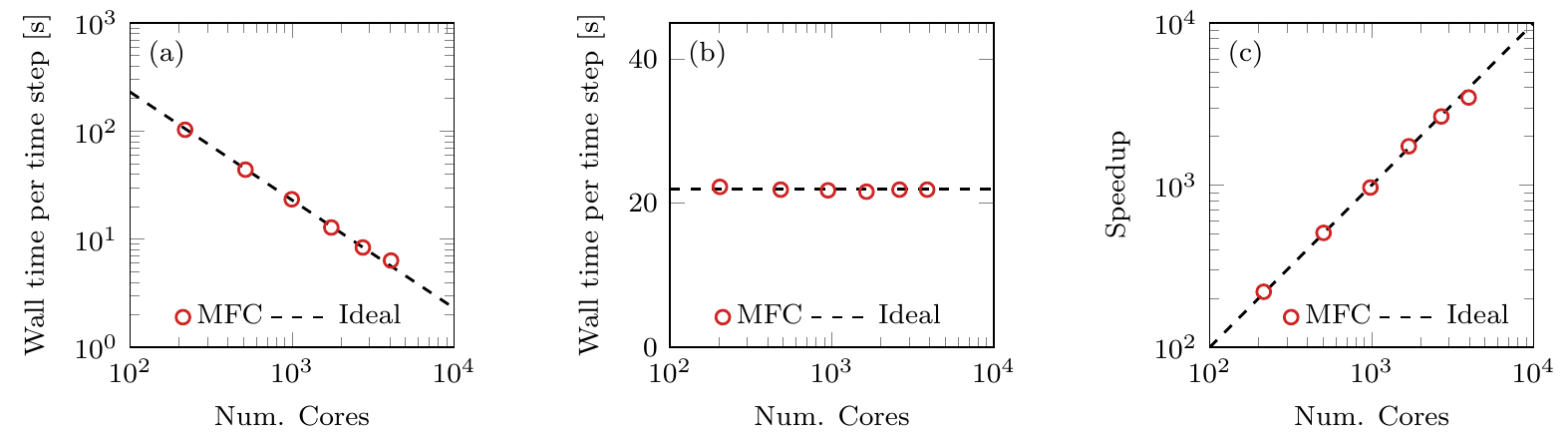}
	\caption{
	MFC performance benchmarks: (a) strong scaling, (b) weak scaling, and (c) speedup tests. 
	}
	\label{f:scaling}
\end{figure}

Figure~\ref{f:scaling} shows parallel performance benchmarks of MFC.  The
strong scaling test of figure~\ref{f:scaling}~(a) measures how the solution
time varies for a fixed problem size as the number of computing cores varies,
the weak scaling test (figure~\ref{f:scaling}~(b)) measures how well the
computational load is balanced across the available cores by measuring solution
time while fixing the grid size distributed to each core and varying the number
of cores (thus changing the overall problem size), and the speedup test
(figure~\ref{f:scaling}~(c)) measures how the solution time increases with
respect to serial computation as the number of cores varies. Thus, speedup is
defined by the ratio of time cost of a parallel simulation with a certain of
cores to a serial computation. The strong scaling and speedup tests are
carried out on a $500^3$ grid, while a constant load of $50^3$ cells per core
is maintained during the weak scaling test. For all tests, MFC performs
very near the ideal threshold until the number of cores is $4096$, at which
point the results deviate slightly from ideal.

\section{Conclusions}\label{s:conclusions}

We presented MFC, an open-source tool capable of simulating multi-component,
multi-phase, and multi-scale flows.  It uses state-of-the-art diffuse-interface
models, coupled with high-order interface-capturing and Riemann solvers, to
represent multi-material dynamics. MFC includes a variety of flow models and
numerical methods, including spatial and temporal orders of accuracy, that are
useful when considering the computational requirements of challenging open
problems. It also includes options for additional physics and modeling
techniques, including a sub-grid ensemble-averaged bubbly flow model. 

We also described the requirements to build MFC and its design. This
included external open-source software libraries that are readily available
online. MFC was divided into three main components that initialize and
simulate the flow, then process the exported simulation data. Each of these
components is modular, and thus can be readily modified by new developers. They
are coupled together via an intuitive input Python script that automatically
generates the required Fortran input files and executes the software component.
The exported simulation files can be readily analyzed or treated via parallel
post-processing.

Finally, we presented a comprehensive set of validations, verifications, and
illustrative examples. Validation was performed via comparisons to
expected bubble dynamics and shock-bubble, shock-droplet, shock-water-cylinder
experiments, while verification was obtained via numerical experiments
involving isentropic and Taylor--Green vortices, as well as advected- and
interface-interaction problems. The capabilities and fidelity of MFC were
also illustrated using challenging studies of
shock--bubble-viscous-vessel-wall interaction in application to shock-wave
lithotripsy and acoustic-bubble-net dynamics in application to feeding humpback
whales. A set of performance benchmarks also showed that MFC was able to
perform near the ideal threshold of parallel computational efficiency.

\section*{Acknowledgments}

The authors are grateful for the suggestions of Dr.\ Benedikt Dorschner when
making MFC open source. This work was supported in part by multiple past
grants from the US National Institute of Health (NIH), the US Office of Naval
Research (ONR), and the US National Science Foundation (NSF), as well as
current NIH grant number 2P01-DK043881 and ONR grant numbers N0014-17-1-2676
and N0014-18-1-2625. The computations presented here utilized the Extreme
Science and Engineering Discovery Environment, which is supported under NSF
grant number CTS120005. K.M.  acknowledges support from the Funai Foundation
for Information Technology via the Overseas Scholarship.

\section*{References}
\bibliographystyle{elsarticle-num-names}
\bibliography{bryngelson_CPC.bib}

\end{document}